**Exploring Convergence in Relation using Association Rules Mining: A Case Study in Collaborative Knowledge Production**

**Jiahe Ling (Corresponding Author)**
Department of Statistics
University of Wisconsin Madison, Madison, WI, 53703
Tel: 608-556-4287; Email: jling9@wisc.edu

**Corey B. Jackson, Ph.D.**
The Information School
University of Wisconsin Madison, Madison, WI, 53703
Tel: 608-263-2900; Email: cbjackson2@wisc.edu

Word Count: 5116 words

ABSTRACT
This study delves into the pivotal role of non-experts in knowledge production on open collaboration platforms. Leveraging Association Rule Mining (ARM), this research unravels the underlying dynamics of collaboration among citizen scientists by quantifying tag associations and scrutinizing their temporal dynamics. The study provides a comprehensive and nuanced understanding of how non-experts collaborate to generate valuable scientific insights. Furthermore, this investigation extends its purview to examine the phenomenon of ideological convergence within online citizen science knowledge production. A novel measurement algorithm based on the Mann-Kendall Trend Test is introduced to accomplish this. This innovative approach sheds light on the dynamics of collaborative knowledge production, revealing both the vast opportunities and challenges inherent in leveraging non-expert contributions for scientific research endeavors. Notably, the study uncovers a robust convergence pattern, employing the newly proposed convergence testing method and the traditional approach based on the stationarity of time series data. This research holds significant implications for understanding the dynamics of online citizen science communities and underscores the crucial role played by non-experts in shaping the scientific landscape. Ultimately, this study contributes significantly to our understanding of online citizen science communities, highlighting their potential to harnecollective intelligence for tackling complex scientific tasks and enriching comprehension of collaborative knowledge production processes in the digital age.

**INTRODUCTION**

Knowledge production, the process of generating knowledge, is essential to many open collaboration platforms. Knowledge production often involves generating, organizing, and disseminating information to enhance existing knowledge. Platforms such as Wikipedia and open-source software development (OSS) communities facilitate collective knowledge production, allowing individuals who often lack formal training to contribute expertise and insights for communal benefit. In Wikipedia, individuals collaboratively write and edit articles; in OSS, they collaborate to write and debug software code.

Another form of collaboration, and the focus of this research, takes place in online citizen science projects. Citizen science describes efforts to engage non-professionals in scientific research through tasks including data collection and data analysis. One example, and the focus of this work, is Gravity Spy. In the Gravity Spy project, volunteers analyze pre-existing datasets and in discussion threads, engage in conversations related to the data, and often produce novel scientific insights.

This participatory model extends beyond content discovery, shaping scientific understanding. However, as citizen science tasks grow more intricate, empirical evidence is needed to justify the involvement of amateur contributors in coordination and communication-intensive projects. The discussions serve as a central focal point for evaluating the abilities of non-expert volunteers to do and engage in science. Questions about non-experts' ability to mimic critical aspects of scientific practice that often involve reasoning, debating, and reaching consensus become central. These discussions showcase the potential to engage with scientific concepts and contribute meaningfully to the broader scientific discourse.

Analytic methods are needed to investigate and evaluate individuals' ability to engage in scientific practice. Association rule learning or mining is a data mining technique to discern relationships or associations between variables within large transaction datasets. AR can unveil patterns or correlations among items predicated on their co-occurrence in transactions or events. AR learning can help uncover patterns and relationships within social interactions and user-generated content within open collaboration platforms, providing valuable insights into user behavior, community dynamics, and content engagement.

This paper has a methodological and applied focus. Methodologically, we propose a novel statistical approach method for examining consensus. Using an ensemble method, we blend association rule learning and the Mann-Kendall trend test to quantify conceptual convergence. In applying the model, we investigate the collaborative knowledge production practices within the Gravity Spy project to understand how knowledge is created and shared among volunteers. We examine association rules weekly and track how metrics (i.e., support and confidence) evolve. This approach enables the examination of the relationship between user-provided tags.

This paper endeavors to investigate the phenomenon of conceptual convergence in online citizen science knowledge production by introducing a novel measurement algorithm. It addresses the question: "How prevalent is conceptual convergence in knowledge production?" Additionally, it aims to compare this new algorithm with traditional methodologies for analyzing convergence.

**LITERATURE REVIEW**

Knowledge production involves generating, organizing, and sharing information to enhance existing knowledge. This process encompasses collecting data, gaining insights, and communicating these insights to others [1]. Knowledge can be explicit (formal, documented) or tacit (personal, experiential). Historically, knowledge production was centralized but has shifted towards decentralization with the advent of the Internet and information technologies [2], [3]. Decentralized production, exemplified by platforms like Wikipedia and OpenStreetMap, helps diversify the population of contributors, leading to a richer knowledge ecosystem. However, challenges such as quality control and sustainability persist [4], [5], [6]. For instance, Wikipedia relies heavily on volunteer contributors to create and edit articles. This openness allows for a broad range of input, but it also leads to potential inaccuracies or biases if not adequately monitored. Solutions proposed include establishing common ground, leveraging user engagement, and infrastructure development. Collaborative tagging, as seen in citizen science projects like Galaxy Zoo and BeeWatch, exemplifies bottom-up categorization and information retrieval, though concerns about content quality remain. Collaborative tagging involves users applying descriptive tags to content, reflecting diverse viewpoints and interpretations, and collectively creating content organization.

This research draws on three bodies of literature to situate the work and propose a method for evaluating conceptual convergence. First, the literature on folksonomies helps us understand the process and outcomes of ideological convergence in knowledge production platforms like those studied in this research. Second, the literature on association rule mining provides background on a technique frequently used to discover relationships in large datasets. This work extends ARM by incorporating evaluations of convergence and stationarity to pinpoint how the relationships derived from an ARM analysis might pinpoint temporal trends related to the tag pair relationships.

**Folksonomies**

Folksonomies, a collaborative tagging system utilized to categorize and organize digital content, is generated by a community of users who assign tags based on personal understanding and perspectives [7]. Research by Makani and Spiteri indicates a decline in unique tags over time within a knowledge management community, suggesting the establishment of a stable and domain-specific vocabulary [8]. Additionally, Kopeinik et al. demonstrated that tag curation significantly contributes to semantic stabilization in group learning projects [9]. Santos-Neto et al. proposed utilizing tagging activity distribution and user interest similarity metrics to enhance navigability in expanding knowledge spaces [10]. However, folksonomies embody a decentralized approach to content categorization, leading to inconsistencies, redundancies, and ambiguities in tag interpretation. For instance, synonymous spelling variants or acronyms challenge observing knowledge production through tagging.

**Association Rule Mining (ARM)**

Association rule mining (ARM) is a widely used technique for uncovering significant patterns in large datasets [11]. Folksonomies, prevalent collaborative tagging systems in social media, serve as valuable data sources for ARM applications. One key use of ARM in folksonomies is exploring patterns and relationships among user-generated tags, revealing insights into the dynamics of folksonomies, community formation, and information diffusion. Researchers have systematically investigated different projections of folksonomy structures to uncover meaningful tag associations [12]. By applying ARM to extensive folksonomy datasets, researchers have demonstrated the utility of discovered rules for various purposes, including tag recommendation, user profiling, resource organization, and community detection [13]. However, the literature suggests that the utilization of ARM in folksonomies remains limited compared to other analytical approaches like tag clouds and tag-based recommendations. Further in-depth analysis of ARM within folksonomies is thus warranted to understand how to leverage this data mining technique in social media contexts. Typically, ARM lacks a temporal dimension. Integrating a temporal component necessitates the adoption of additional analytical methods to effectively interpret the data.

**Convergence and Stationarity**

The term "convergence" carries varied meanings across different contexts. In economics, convergence typically denotes the diminishing disparities or distinctions among regions, entities, or variables over time [14]. In some other fields, the "convergence" of a time series is also used interchangeably with the term "stationarity," which entails the constancy of statistical properties over time [15]. However, this paper defines convergence more strictly: it signifies a time series approaching a specific value through continuous variation reduction as time advances.

In collaborative learning, convergence is not determined through statistical testing but conceptualized as an ideological alignment among individuals. Experiments and surveys are typically conducted to discern the factors influencing convergence and ascertain whether individuals reach an agreement [16].

In summary, no quantitative (statistical) methodology currently exists to objectively measure the convergence of individuals toward a particular idea in collaborative learning. Existing methods can only detect if a time series diminishes disparities or achieves stationarity relative to another. This underscores the significance of this research, which aims to quantify the learning patterns of a collaborative tagging system using ARM as a time series. Subsequently, this paper proposes a methodology to identify the point in time at which the series approaches a specific value through continuous reduction of variation as time progresses.

**Gravity Spy**

The research focuses on Gravity Spy, an online citizen science project where volunteers collaborate with researchers to identify and categorize noise signals in gravitational wave data. Volunteers assist by classifying datasets into known categories and developing new ones. They use a classification interface with a Field Guide for reference and metadata options. As volunteers gain expertise, they progress to more advanced workflows. Additionally, volunteers engage in discussions on the Gravity Spy Talk platform, particularly regarding unidentified glitches, which may signify new categories. Volunteers can propose new glitch classes by submitting detailed proposals, which LIGO scientists evaluate. These proposals include images, descriptions, and proposed names for the new glitch classes. The process aims to formalize the creation of new glitch categories based on volunteer input. An example of the Gravity Spy classification interface is shown in **Figure 1**.

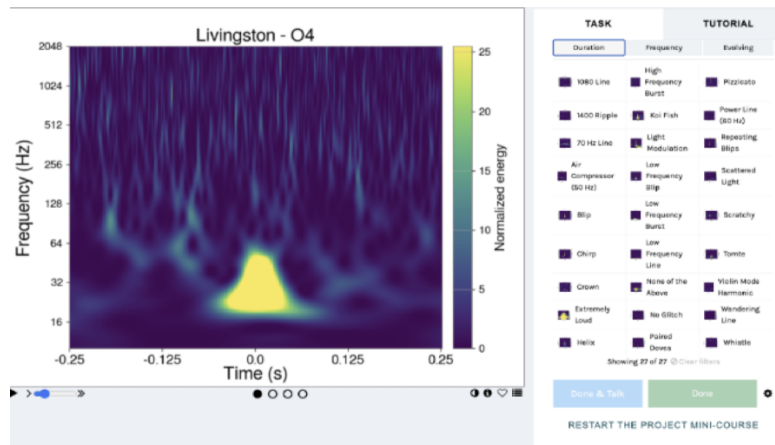

**Figure 1 The Gravity Spy classification interface: an image of a glitch on the left, and the list of classification options on the right.**

It is important to note that the platform consists of various discussion boards, each containing multiple discussions. Within each discussion, numerous hashtags could be proposed. Comments posted on boards titled "Notes" were included to streamline the analysis. Unlike other discussion boards, "Notes" boards are associated with image subjects, suggesting that hashtags on these boards likely relate to new glitch class proposals rather than general discussions.

**DATA**

This study gathered digital trace data encompassing all comments posted by volunteers on Zooniverse between October 2016 and July 2023. Emphasis was placed on hashtags as they are the primary vehicles for linking image objects across the platform. Each entry in the hashtag dataset included a timestamp indicating the posting time, along with metadata such as board_id, discussion_id, the hashtag itself, and the username of the volunteer who posted the comment.

The original dataset comprises 144,593 observations in total. During the analysis process, this study utilized data sourced from the discussion threads of proposal tags, which this study refers to as "seed tags." Each discussion thread associated with glitch proposals contained an average of 4.41 (with a standard deviation of 6.31 and a median of 3) seed tags. As the number of tags in these discussion boards reflects the final curation, this study delved into the discussions centered around image subjects related to these seed tags to gain deeper insights into the evolution of tagging dynamics. This study extracted the seed tags for every proposal and then identified the discussion IDs of comments containing each seed tag. This discussion threads serve as platforms where volunteers engage in conversations regarding the presence of noise signals, often including additional tags beyond those initially submitted with the proposals. This broad approach was deliberately chosen due to uncertainties surrounding the current relationships among tags in the project.

Implementing this search strategy yielded a dataset comprising 61,657 comments (representing 42% of all comments) containing a total of 78,803 tags, out of which 4,219 were unique. Less than 12% of volunteers contribute comments in Gravity Spy, with 53% (N = 1,749) of those volunteers included in our dataset. On average, each proposal contained 360 tags (with a standard deviation of 481 and a median of 80). The disparity between the average number of seed tags and those identified through our search indicates a challenge for volunteers in fully expressing the complete range of tags and their relationships. Each proposal involved an average of 167 volunteers (with a standard deviation of 313 and a median of 39).

|  | LHS | RHS | support | confidence | count | week |
|---|---|---|---|---|---|---|
| 33539 | {#helix} | {#possiblenewglitch} | 0.0021 | 0.3333 | 1 | 42 |
| 35026 | {#helix} | {#possiblenewglitch} | 0.002 | 0.3333 | 1 | 43 |
| 36615 | {#helix} | {#possiblenewglitch} | 0.002 | 0.3333 | 1 | 44 |
| 38220 | {#helix} | {#possiblenewglitch} | 0.0019 | 0.3333 | 1 | 45 |
| 39829 | {#helix} | {#possiblenewglitch} | 0.0019 | 0.3333 | 1 | 46 |
| 41431 | {#helix} | {#possiblenewglitch} | 0.0019 | 0.3333 | 1 | 47 |
| 43051 | {#helix} | {#possiblenewglitch} | 0.0018 | 0.3333 | 1 | 48 |
| 44672 | {#helix} | {#possiblenewglitch} | 0.0018 | 0.3333 | 1 | 49 |
| 46304 | {#helix} | {#possiblenewglitch} | 0.0018 | 0.3333 | 1 | 50 |
| 47934 | {#helix} | {#possiblenewglitch} | 0.0017 | 0.3333 | 1 | 51 |

**FIGURE 2. A Screenshot of a Subset of the Data.**

**METHODOLOGY**

This research utilized association rule mining, a data mining technique to uncover dataset relationships, patterns, and dependencies. Association rules consist of an antecedent (or left-hand side) and a consequent (or right-hand side), with the antecedent identifying significant relationships between items and the consequent representing another item likely to co-occur. This approach has wide-ranging applications, including market basket analysis, customer behavior analysis, and recommendation systems [1], [2], [3].

In this study, tags were treated as items and dependencies as the co-occurrence of tags. The output of association rules provided insights into the frequency and strength of dependencies, measured through ARM values *support*, *confidence*, and *lift*. Support gauges how frequently a specific item set occurs, confidence quantifies the conditional probability of the consequent given the antecedent, and lift compares the likelihood of the consequent occurring given the antecedent.

This analysis proceeded in three steps: firstly, capturing the discussion_ids of seed tags; secondly, creating a subset of comments with the same discussion_id as the seed tags; and finally, computing association rules with support and confidence thresholds set at 0.001 to ensure the inclusion of both strong and weak associations. These rules were calculated on an aggregated weekly basis, offering snapshots of association patterns within the dataset over time.

Following association rule mining, the dataset expands to encompass 311,114 observations, including 42 proposals and 610 tag pairs. The dataset included weekly measures of support for each tag pair relationship. Since the output shows support over many periods, additional analytic methods are required to better understand conceptual convergence. The research relies on trend detection to evaluate convergence. In the space below, we describe the approach in more detail.

**Mann-Kendall Test**

The Mann-Kendall test, a non-parametric statistical method, is employed to identify monotonic trends within time series datasets. Unlike parametric tests, it does not rely on assumptions of normality or linearity, rendering it robust across diverse data types. Monotonic trends signify consistent increases or decreases in a variable over time without necessitating linear patterns. In the context of this test, the null hypothesis suggests an absence of monotonic trends within the data, while the alternative hypotheses encompass scenarios of upward, downward, or any form of monotonic trend.
In this study, the hypotheses are formulated as follows:

$$H_0: \text{No monotonic trend}$$
$$H_a: \text{Downward monotonic trend}$$

First, Arrange the data sequentially based on the collection order over time, denoting the measurements as $x_1, x_2, \ldots, x_n$, where $x_i$ represents the measurement obtained at time $i$ for $i = 1, 2, \ldots, n$. Then, calculate the sign of all possible differences $x_j - x_k$, where $j > k$. The number $S$ is calculated through the formula in (1) where $sgn(x)$ is defined in (2):

$$S = \sum_{k=1}^{n-1} \sum_{j=k+1}^{n} sgn(x_j - x_k) \qquad (1)$$

$$sgn(x) = \begin{cases} 1 & \text{if } x > 0 \\ 0 & \text{if } x = 0 \\ -1 & \text{if } x < 0 \end{cases} \qquad (2)$$

If $S$ is a positive number, it indicates that observations obtained later tend to be larger than observations made earlier. Conversely, if $S$ is a negative number, observations made later tend to be smaller than observations made earlier. The variance could be calculated through the formula in (3) where $n$ is the number of data points in the time series, $m$ is the number of tied groups in the data, $t_p$ is the number of data points in the $p$th tied group, and $q_p$ is the number of data points minus the number of tied groups:

$$\text{Var}(S) = \frac{n(n-1)(2n+5) - \sum_{p=1}^{m} q_p(t_p-1)(t_p-2)}{18} \qquad (3)$$

Finally, the test statistic $Z_{MK}$ is calculated through the formula in (4):

$$Z_{MK} = \begin{cases} \frac{S-1}{\sqrt{\text{Var}(S)}} & \text{if } S > 0 \\ 0 & \text{if } S = 0 \\ \frac{S+1}{\sqrt{\text{Var}(S)}} & \text{if } S < 0 \end{cases} \qquad (4)$$

The $H_0$ is rejected if, at the significance level $\alpha$, $Z_{MK} \leq -Z_{1-\alpha}$, and the trend could be considered as Downward monotonic trend. In this study, the significance level $\alpha$ is established at 0.05.

**Test for Convergence**

While convergence is often equated with stationarity in various studies, its strict definition entails a time series gradually nearing a specific value with a continuous reduction in variation over time.

Some research endeavors to pinpoint this convergence by establishing a threshold and identifying the time point where the series fluctuates within this threshold. However, this approach encounters challenges as the optimal threshold can vary across different data types, posing a bottleneck to the methodology. Additionally, there is a risk of falsely claiming convergence, as the variance may not consistently diminish over time.

Thus, rather than pinpointing the convergence itself, this paper proposes an algorithm to detect the initiation of convergence, defined as the point where the time series' variance begins to decrease consistently. This approach aims to mitigate the uncertainty in concluding convergence by focusing on the observable onset of reduced variation in the time series.

```
Algorithm 1: FindConvergeStartPoint
  Data: A time series L
  Result: The time point t where the convergence begins in L
  Input  : L, α
  Output: t
  t ← ∞;
  N ← length(L);
  // Step 1:  Calculate the Standard Deviations
  for i ← 1 to N − 1 do
  |   S[i] ← var(L[i : N]);
  end
  // Step 2:  Mann-Kendall Test, H_a : Downward monotonic trend
  for i ← 1 to N − 3 do
  |   p ← Mann-Kendall(S[i : N − 1]);
  |   τ ← Mann-Kendall(S[i : N − 1]);
  |   if p < α and τ < 0 then
  |   |   t ← i;
  |   |   return i;
  |   end
  end
  return t;
```

The *FindConvergeStartPoint* algorithm is designed to precisely estimate when convergence begins within a given time series, a crucial aspect in various analytical contexts. Its functionality relies on two primary inputs: the time series data and a user-defined significance level. Operating on these inputs, the algorithm systematically processes the time series data to determine the point at which convergence initiates.

The algorithm unfolds in a series of steps, each geared towards identifying the onset of convergence with rigor and efficiency. Initially, it initializes a variable to a theoretically infinite value, symbolizing the absence of any observed convergence point at the outset. Subsequently, it calculates the standard deviations of consecutive subsets of the time series data, starting from each data point and extending to the end of the series. This step captures the variability inherent in different time series segments, laying the groundwork for subsequent analysis.

Following the standard deviation calculations, the algorithm conducts a Mann-Kendall test on these computed standard deviations. The Mann-Kendall test, discussed in the previous section, evaluates whether a statistically significant downward trend exists in the standard deviations. Such a trend, if present, signifies a consistent reduction in variability over time, a hallmark of convergence in many contexts.

If the Mann-Kendall test yields a p-value below the specified significance level, indicative of a significant downward trend in standard deviations, the algorithm returns the time at which this trend commences. This identified time point is the estimated start of convergence within the time series data.

The *FindConvergeStartPoint* algorithm offers a methodical approach to discerning the incipient convergence phase within a time series. Leveraging statistical tests and user-defined significance levels facilitates a nuanced understanding of the evolving dynamics present in the data, empowering researchers to make informed interpretations and decisions based on the identified convergence points.

The program complexity of the *FindConvergeStartPoint* algorithm can be assessed in terms of time and space requirements. Regarding time complexity, the algorithm entails two primary loops, each iterating over the length of the input time series, denoted as $N$. The first loop calculates standard deviations for subsets of the time series, resulting in a time complexity of $O(N^2)$, as it iterates over each data point and computes the standard deviation for subsequent subsets. The second loop conducts the

Mann-Kendall test, also iterating over the time series length, with each iteration involving a calculation based on the Mann-Kendall test, typically performed in linear time. Therefore, the overall time complexity of the algorithm remains $O(N^2)$.

Concerning space complexity, the algorithm necessitates additional space to store the standard deviations computed for each subset of the time series, resulting in a linear space complexity of $O(N)$. Moreover, a few auxiliary variables, such as $t$, require constant space.

**RESULTS AND DISCUSSION**

**Start Point of Convergence**

This study observed that 99.8% of support metric time series converge emphasizes the consistency of these metrics over time, signifying a robust tendency for tag pairs to stabilize in their relationships before the proposal submission deadline. This high convergence rate implies a predictable pattern in the behavior of tag pair relationships, instilling confidence in their reliability as indicators of collaborative dynamics.

Moreover, the average convergence start point occurring approximately 2.3 weeks with a median of 1 week and standard deviation of 5.26 weeks before the proposal submission. This timeframe allows volunteers to refine collaborative strategies up until the submission point, potentially optimizing research directions and enhancing the efficacy of joint efforts. However, the data still shows a large variation between different tag groups.

The early onset of convergence also implies a period of adaptation and refinement in tag pair relationships during the weeks leading up to proposal submission. This phase likely involves iterative discussions, resource allocation, and alignment of objectives, fostering a conducive environment for productive collaboration. By identifying this trend, researchers and stakeholders gain valuable insights into the dynamics of collaborative relationships, enabling them to anticipate and navigate the proposal submission process with greater clarity and efficiency.

Overall, the high convergence rate and early stabilization of tag pair relationships highlight the reliability and predictability of collaborative dynamics, offering valuable guidance for effective research collaboration and proposal development.

**Figure 3** presents a support metric time series that has remained unconverged to visually represent the proposed algorithm. Upon examination, no discernible pattern in variance or trend is evident throughout the entire time frame. This lack of convergence is visually depicted, emphasizing the absence of stabilization or consistency in the relationship dynamics captured by the support metric.

In contrast, **Figure 4** illustrates a support metric time series that has converged at 59 weeks before the proposal submission. A clear pattern emerges in decreased variation across the entire time series, accompanied by a noticeable stabilization of values. This visual depiction portrays the convergence of the support metric, highlighting the consistent and stable nature of the relationship dynamics captured within the data.

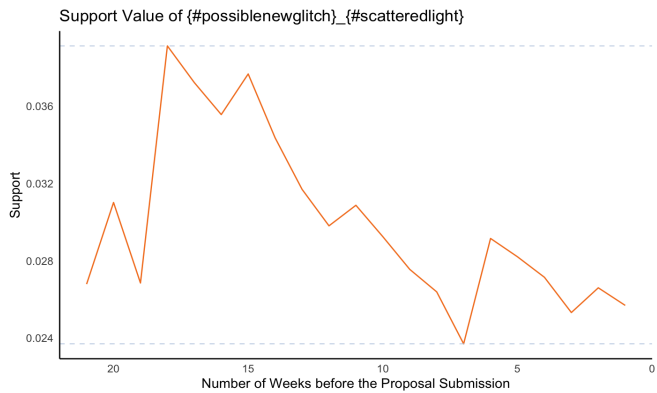

**FIGURE 3.** The trend in association rule support for {#possiblenewglitch} -> {#scatteredlight} pair showing no trend of convergence.

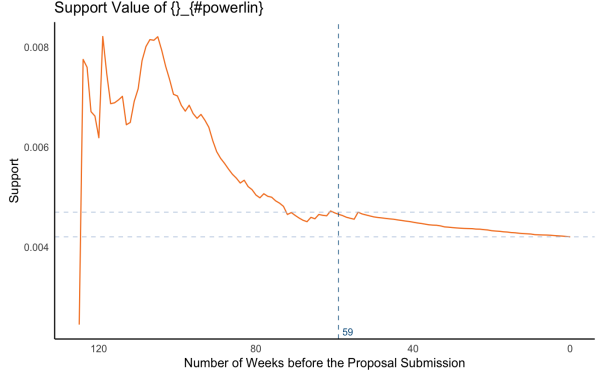

**FIGURE 4.** The trend in association rule support for { } -> {#powerlin} pair shows a trend of convergence at 59 weeks before proposal submission.

**Analysis of Stationarity**

In the context of time series analysis, a stationary time series is one where statistical property such as mean, variance and autocorrelation structure remain constant over time. This study assesses the stationarity of the "support" metric derived from association rule analysis of tag pairs. To accomplish this, the Augmented Dickey-Fuller test, a widely used tool for testing the stationarity of a time series, is employed.

The methodology for pinpointing the time stationarity is achieved and involves stepwise analysis of the support metric's time series. The analysis begins from the initial appearance of the tag pair and proceeds incrementally until the proposal submission deadline. Initially, the Augmented Dickey-Fuller test is applied to evaluate the stationarity of the time series spanning this entire duration [17].

Should the initial analysis reveal non-stationarity, the focus narrows to the period starting from the second week after the tag pair's introduction and extends up to the proposal submission deadline. This iterative process continues, with each subsequent analysis commencing one week later than the previous one until the penultimate week before the proposal submission deadline.

The corresponding week number is recorded throughout this iterative procedure if the Augmented Dickey-Fuller test indicates a stationary time series for the support metric. Conversely, if the test consistently fails to identify stationarity across all iterations, it is deduced that the support for the tag pair does not attain stationarity within the observed timeframe. This approach ensures a detailed exploration of the dynamics of the support metric and provides insights into its behavior over time.

The analysis unveils that 74% of tag pairs display stationarity before the proposal submission deadline. Among these stationary pairs, a detailed examination reveals that the average timeframe for becoming stationarity is 100 weeks before the proposal submission, with the median convergence taking place at 96 weeks. These findings imply that most tag pairs stabilize their relationships well before the formal proposal submission process.

A case study centered on {}_{#powerlin} was conducted to compare the newly proposed methodology in this paper with traditional stationarity analysis. According to the new methodology by detecting the start point of convergence, {}_{#powerlin} began to converge 59 weeks prior to the submission of the proposal. In contrast, the traditional stationarity analysis suggests that {}_{#powerlin} has not achieved stationarity at any point in time.

Another case study which mainly focused on the tag pair {#helix}_{#possiblenewglitch} is conducted to demonstrate the concept of convergence. Employing the methodology outlined earlier, the convergence point is observed in the 181st week preceding the proposal submission. **Figure 5** depicts the comprehensive trend of the support metric spanning from 200 days before the proposal submission to the actual day of proposal submission.

In **Figure 5**, the support value post-convergence exhibits fluctuations within the range of 0.003585347 to 0.007125891, with a variance of 2.667543e-07. It becomes evident that the support value remains consistently stable around a specific value at the 181st week before the proposal submission. This detailed examination clearly illustrates how stationarity manifests within tag pair relationships.

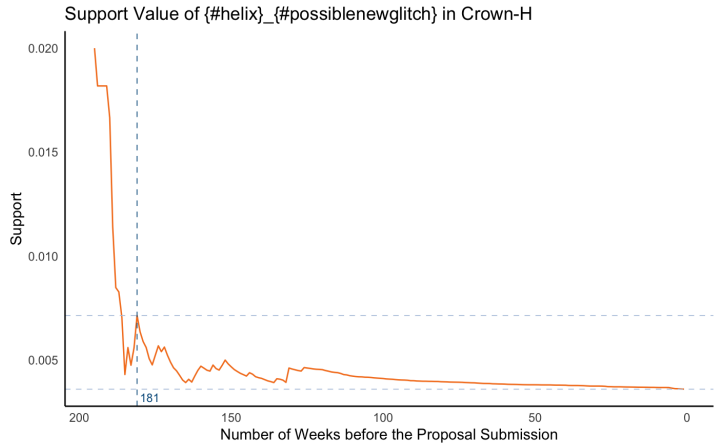

**FIGURE 5. The trend in association rule support for {#helix} -> {#possiblenewglitch} pair showing a trend of stationarity at 181 weeks before proposal submission.**

A discernible contrast between **Figure 4** and **Figure 5** reveals distinct characteristics of stationarity and convergence. Stationarity encompasses a phase where values may exhibit fluctuations, whereas convergence entails a gradual stabilization toward a specific value. While **Figure 4** depicts the transition towards convergence with diminishing variability and an increasingly stable trend, **Figure 5** illustrates a lack of convergence characterized by ongoing fluctuations without a discernible stabilization pattern. This comparison underscores the significance of identifying convergence points, as they signify a pivotal moment where the time series dynamics become more predictable and stable.

## CONCLUSION AND LIMITATIONS

In conclusion, this research offers valuable insights into the dynamics of tag pair relationships in the context of proposal submissions. Through analysis of the support metric time series, it becomes evident that most tag pairs exhibit convergence well before the proposal submission deadline, indicating a propensity for stability in collaborative dynamics. The observation of high convergence rates and early stabilization points underscores the predictability and reliability of tag pair relationships, offering researchers and stakeholders valuable guidance in collaborative endeavors. Moreover, the visual representations provided by the support metric time series offer clear illustrations of convergence patterns, highlighting the importance of identifying and understanding these critical junctures in collaborative dynamics. By shedding light on the processes leading up to proposal submissions, this research contributes to a deeper understanding of collaborative dynamics in research environments, ultimately facilitating more effective collaboration and proposal development strategies.

While the algorithm presented offers a structured method for pinpointing the convergence start point within a time series, it is imperative to acknowledge its inherent limitations. The algorithm primarily leans on the Mann-Kendall test to discern downward monotonic trends in the time series. While widely utilized for trend detection, this test may not be universally applicable, potentially yielding inaccurate outcomes under certain circumstances such as non-linear trends or outlier presence. Additionally, the algorithm's efficacy is contingent upon parameter selection, including factors like the length of the sliding window for standard deviation calculations and the step size for traversing the time series. Poor parameter choices could compromise the precision and reliability of the identified convergence start point. Furthermore, the Mann-Kendall test's conservative nature may overlook conspicuous trends, further complicating accurate trend detection. Despite providing a structured framework for convergence analysis, the algorithm may lack robustness and scalability when confronted with extensive or intricate datasets. The algorithm's iterative nature and computational demands could render it impractical for large-scale or high-dimensional data analysis, posing significant challenges for real-world applications.

While this study offers valuable insights, it is important to acknowledge its limitations. Firstly, the dataset may suffer from sampling bias, potentially limiting the generalizability of the findings to a broader population or context. Additionally, there may be instances of incomplete or missing data, which could introduce gaps in the analysis. The quality and accuracy of the extracted tags from discussions may vary, influenced by inconsistencies, misspellings, or ambiguities introduced by volunteers. Moreover, volunteer behavior within the discussions could introduce biases, with certain individuals being more active or influential than others. It's also essential to recognize that correlation does not imply causation, and further investigation would be needed to establish causal relationships. The study may not fully capture the evolution of tagging practices over time, and the analysis may be constrained by the scope and depth of the research. Furthermore, the findings may be context-specific and may not be applicable to other domains or settings.

The tags employed in the project serve diverse purposes. For instance, the tag #possiblenewglitch is frequently used in many proposals to suggest the potential discovery of a new phenomenon within the data. Similarly, #helix is commonly associated with a characteristic of many subjects in images and appears in numerous proposals. In contrast, #1800ripple is often used to tentatively name a new glitch class. While the convergence of tags like #helix and #possiblenewglitch may not yield significant insights, the use of #1800ripple could be more intriguing. Consequently, not all tag combinations hold the same value. Future research could involve differentiating tags based on their broader usage within the project, potentially excluding pairs that include #possiblenewglitch to refine the analysis.
By acknowledging these limitations, the study can provide a more nuanced understanding of its findings and guide future research efforts to address these gaps.

**ACKNOWLEDGMENT**
This publication uses data generated via the Zooniverse.org platform, the development of which is funded by generous support, including a Global Impact Award from Google, and by a grant from the Alfred P. Sloan Foundation. J. Ling wishes to express profound gratitude to Dr. Corey B. Jackson, the honor thesis advisor, whose unwavering support, insightful guidance, and profound expertise have been invaluable throughout this journey. J. Ling is deeply appreciative of Carmela Diosan, the Academic Advising Manager, whose assistance with administrative tasks has been indispensable. Heartfelt thanks are extended to J. Ling's beloved family for their boundless love, encouragement, and sacrifices, which have been the cornerstone of strength and motivation. Equally, gratitude is extended to friends and peers whose unwavering encouragement and support have been a constant source of inspiration. Furthermore, heartfelt acknowledgment is given to the University of Wisconsin-Madison, Department of Statistics, and the University of Wisconsin-Madison, The Information School, for their unwavering commitment to providing the necessary resources and conducive environment for this research endeavor. J. Ling is also deeply grateful to the Purdue University Summer Undergraduate Research Fellowship (SURF) program for providing invaluable opportunities for research exploration and growth. Special appreciation is extended to Dr. Konstantina Gkritza and Dr. Xiaodong Qian, J. Ling's first research mentors. Their mentorship, guidance, and unwavering belief in their potential ignited the spark that set J. Ling's research journey in motion. Their dedication to fostering an environment of intellectual curiosity and scholarly development laid the groundwork for a lifelong pursuit of knowledge and inquiry. Each of these contributions, whether through guidance, support, or resources, has played an integral role in the completion of this thesis. For these countless gestures of kindness, encouragement, and support, J. Ling is genuinely thankful. This thesis stands as a testament to the collective effort and collaboration of all involved, and their contributions will forever be remembered with deep appreciation and gratitude.